\documentclass[aps,nofootinbib,prd,superscriptaddress,tightenlines,notitlepage,twocolumn,showpacs,floatfix]{revtex4-1}

\usepackage[colorlinks]{hyperref}
\usepackage{tabularx}
\usepackage{bm}
\usepackage{graphicx}
\usepackage{amssymb,bm,tensor}
\usepackage{xcolor}
\usepackage{amsmath}
\usepackage[varg]{txfonts}
\usepackage{enumerate}
\usepackage[normalem]{ulem}
\usepackage{bm}
\usepackage[OT1]{fontenc}
\usepackage{float} 

\usepackage{scalerel,stackengine}

\newcommand{\hwsout}{\textcolor{cyan}{$\mathcal{HT}$: }\bgroup\markoverwith{\textcolor{cyan}{\rule[.5ex]{2pt}{2.5pt}}}\ULon}

\begin{document}


\title{Constraints on charged black holes from merger-ringdown signals in GWTC-3\\
 and prospects for the Einstein Telescope}

\author{Hua-Peng Gu}
\author{Hai-Tian Wang}
\email[Corresponding author: ]{wanght@pku.edu.cn}
\affiliation{Department of Astronomy, School of Physics, Peking University, Beijing 100871, China}
\affiliation{Kavli Institute for Astronomy and Astrophysics, Peking University, Beijing 100871, China}
\author{Lijing Shao}
\email[Corresponding author: ]{lshao@pku.edu.cn}
\affiliation{Kavli Institute for Astronomy and Astrophysics, Peking University, Beijing 100871, China}
\affiliation{National Astronomical Observatories, Chinese Academy of Sciences, Beijing 100012, China}

\date{\today}

\begin{abstract}
Whether astrophysical black holes (BHs) can have charge is a question to be
addressed by observations.  In the era of gravitational wave (GW) astronomy, one
can constrain the charge of a merged BH remnant using the merger-ringdown signal
of the GW data. Extending earlier studies, we analyze five GW events in GWTC-3,
assuming Kerr-Newman BHs.  Our results show no strong evidence for a charged BH,
and give a limit on the charge-to-mass-ratio $Q<0.37$ at $90\%$ credible level
(CL).  Due to the charge-spin degeneracy in the waveform and the limited
signal-to-noise ratios (SNRs), it is challenging for LIGO/Virgo/KAGRA
observations to provide better constraints. We further simulate data for the
Einstein Telescope (ET), where SNRs can be as large as $\sim270$ in the ringdown
signal.  These simulated events allow us to consider the 220, 221, and 330
ringdown modes altogether, which can help break the charge-spin degeneracy.
{The analysis of a simulated GW150914-like signal shows that ET can
improve  the constraints on the charge-to-mass-ratio to $Q \lesssim 0.2$ at
$90\%$ CL with one ringdown signal.}
\end{abstract}

\maketitle


\section{\label{sec:Introduction}Introduction}

The existence of black holes (BHs), as predicted by Einstein's General
Relativity (GR), has been  substantiated through a series of observations
\citep{BH1,BH2}. These astronomical observations not only empirically elucidate
the physical attributes of BHs but also serve as a robust platform for
validating different BH models and theories of gravity. With the groundbreaking
detection of the gravitational wave (GW) event GW150914 by the LIGO/Virgo
Collaboration \cite{Abbott2016}, GW astronomy has emerged as a potent tool for
probing the properties of BHs. Recent observations by the LIGO/Virgo/KAGRA (LVK)
Collaboration have achieved signal-to-noise ratios (SNRs) as high as $26$ for
binary BH (BBH) merger events \cite{Abbott2016b}. Looking ahead, next-generation
GW observatories like the Einstein Telescope (ET) \cite{Punturo2010} are
anticipated to reach SNRs exceeding $200$ for BBH mergers, thereby enabling even
more precise measurements of BH properties \cite{ET2, Sathyaprakash:2019yqt,
Kalogera:2021bya}.

In the context of GR, the most general solution for a stationary, asymptotically
flat BH is the Kerr-Newman (KN) BH, characterized by mass, spin, and charge
\cite{Kerr, KN}. However, astrophysical BHs are generally considered to be
Kerr-like, as they are expected to be electrically neutral due to de-charging
effects from their surrounding environment. To illustrate this, {consider a toy
model of a BH with mass \( M \) and charge \( +q_M \), surrounded by particles of
mass \( m \) and charge \( \pm q_m \)}. If the BH charge is sufficiently large
such that, with \( G=1 \) in Gaussian units,
\begin{equation}
    {q_M q_m > Mm},
\end{equation}
the electromagnetic force will dominate over gravity, causing the BH to attract
only negatively charged particles. Consequently, the BH will become neutralized
on a timescale far shorter than the timescale associated with GWs
\citep{Gibbons1975, neglectcharge, KNneutral}. Moreover, even in the vacuum,
processes like vacuum polarization and pair production contribute to the
neutralization of the BH \cite{BZ1977}. The upper limit for the charge-to-mass
ratio of a BH has been estimated to be
\begin{equation}
    {Q \equiv {q_M} / {M} \lesssim 10^{-5}\frac{M}{M_{\odot}}},
\end{equation}
thereby reinforcing the Kerr hypothesis for BHs \cite{Cardoso2016}.

Empirical evidence lends credence to the theoretical premises discussed above.
For instance, observations of the bremsstrahlung surface brightness decay in the
Galactic central BH, Sgr A*, suggest that its charge-to-mass ratio is less than
\(10^{-18}\) \citep{galactic1,galactic2}. {Since the discharge mechanisms mentioned above apply for all BHs regardless of their mass, all the astrophysical BHs should have negligible charge.} It is worth noting that the term
``charge" in this context refers exclusively to the {electromagnetic charge. Before we only talked about the electric charge, while the BHs may carry magnetic charge by primordial magnetic monopoles \citep{1984ARNPS..34..461P}. The electric and magnetic charges are indistinguishable for a perturbed BH in vacuum \citep{KNinspiral,2023arXiv230210942P,2023arXiv230615751D}.} However,
theories extending beyond the Standard Model or GR permit BHs to carry different
kinds of charges (see e.g. Ref.~\cite{Xu:2022frb}). The foregoing analysis, for
example, relies on the high charge-to-mass ratio of electrons. In contrast, the
minicharged dark matter model, predicated on an additional hidden \(U(1)\) gauge
field, could give rise to particles with lower charge-to-mass ratios, thereby
making charged BHs less constrained \citep{Cardoso2016,dm}. Moreover, certain
modified gravity theories also admit the possibility of BHs carrying specific
kinds of charges, which would influence the metric and yield observable effects.
As such, direct observations of the metric offer a model-independent avenue for
testing these theories.

Nonetheless, conventional methods for constraining the BH charge via metric
measurements in its vicinity tend to yield rather weak limits. For instance, the
Event Horizon Telescope's recent imaging of Sgr A* has established an upper
bound of \(0.72\) for the charge-to-mass ratio at a \(68\%\) credible level (CL)
\cite{EHT, EventHorizonTelescope:2022wkp, EventHorizonTelescope:2022xqj,Vagnozzi:2022moj}. More
stringent constraints can be garnered from GW detectors, particularly in the
context of BBH mergers. In such events, the GW signal during both the inspiral
and ringdown phases serves as a probe of  BH properties pre- and post-merger,
respectively. Our analysis is primarily concerned with the latter. At the
ringdown stage, the GW signal manifests as a composite of damped sinusoids,
known as quasi-normal modes (QNMs) \cite{qnm}. Both the oscillation frequency
and damping time of these QNMs are dependent on the attributes of the resultant
merged BH, including its charge. By extracting QNM frequencies from the ringdown
data, one can deduce the parameters characterizing the QNM spectrum---a
technique commonly referred to as BH spectroscopy \citep{BHspec,BHspec2}.

There are {three primary challenges} associated with constraining the
charge of the resultant BH using GW detectors. The first challenge pertains to
the computation of QNMs for KN BHs. The interplay between electromagnetic and
gravitational perturbations complicates {the analytical derivation of
QNM frequencies for KN BHs  \citep{Kokkotas1993,Berti2005}. Initial perturbative
solutions for KN BH QNMs were first obtained by \citet{pani1,pani2} in the
slow-rotation limit,  by \citet{KNqnm4} under the weakly-charged condition, and
by \citet{Zimmerman2016} for extremal KN BHs. \citet{WHT2021} later employed
geodesic correspondence to approximate KN QNMs in the eikonal limit. This
approximation constrains the charge-to-mass ratio of the BH remnant in GW150914
to be less than \(0.38\) at \(90\%\) CL, {while this result made an ansatz
for QNMs at high values of the charge.}} Moreover, numerical solutions,
especially for certain dominant QNMs have been calculated by
\citet{KNqnm1,KNqnm2,KNqnm3}. For the inspiral stage, however, a complete KN
solution remains elusive. Existing studies have primarily focused on numerical
simulations at some certain limits, such as for low charge-to-mass ratios
\citep{WHT2,KNinspiral,KNinspiral2,KNinspiral3,KNinspiral4}. Consequently, a
comprehensive inspiral-merger-ringdown (IMR) signal analysis for KN BHs is
currently unfeasible \cite{IMR}. The second challenge arises from the limited
sensitivity of existing GW detectors. Even for high-confidence events like
GW150914, {the SNRs for postmerger data  barely exceed \(10\) \cite{q}}.
This limitation hampers our ability to extract higher-order QNMs from the
ringdown data. When only dominant modes are considered, the charge and spin
parameters become strongly degenerate, rendering precise constraints on the BH
remnant's charge impracticable.  {The third challenge pertains to data
analysis of ringdown signals, including the determination of the start time of
the ringdown signal.  In this work, following previous studies
\citep{WHT2021,q,GWTC3test, Wang:2023mst}, we assume that it starts from the
peak amplitude when including the first overtone mode.}

The GW Transient Catalog (GWTC) provides a compendium of GW events observed by
the LVK Collaboration \citep{GWTC1,GWTC2,GWTC3}. This dataset has been
extensively employed in various BH analyses, including tests of GR and its
modifications. Prior analyses have rigorously measured parameters such as the
mass and spin of the remnant BHs, providing a basis for validating subsequent
model-dependent studies \citep{GWTC1test,GWTC2test,GWTC3test}. In testing GR,
the working model is typically Kerr-like, and to date, no significant deviations
from GR have been observed. \citet{q} leveraged the merger-ringdown signals from
the GWTC-2 dataset \cite{GWTC2} to constrain the charge of remnant BHs.
{Utilizing numerical solutions for dominant QNMs from \citet{KNqnm2,KNqnm3},
they performed} a KN BH analysis and derived the charge-spin distribution for
several high-confidence GW events. A prior based on previous IMR analyses of
Kerr BHs was also applied to further constrain the remnant BH charge. Their most
stringent constraint on the charge-to-mass ratio is less than \(0.33\) at
\(90\%\) CL for GW150914. Building upon this work, we employ GWTC-3 data
\cite{GWTC3} to scrutinize the charge of remnant BHs in five high-credibility GW
events. Our results show negative Bayes factors between the KN and Kerr model.
Consequently, our analysis does not strongly support the existence of charged BH
remnants. Due to the low SNRs in ringdown data, coupled with charge-spin
degeneracy, the error margins in our analysis are substantial even with the
improved noise levels in current LVK observations. Although our constraints are
substantially weaker than those derived from electromagnetic observations,
{which can be on the order of \(Q \sim 10^{-18}\)}, our approach is
complementary, and offers the advantage of being model-independent.

Moreover, to assess the potential for detecting BH charges with the future GW
detector ET, we conduct analyses using simulated GW data. Designed to achieve a
tenfold increase in sensitivity \cite{Punturo2010}, ET has not yet been
constructed, but simulated data analyses can offer valuable insights into future
constraints on remnant BH charges using merger-ringdown signals. Enhanced
detector sensitivity not only reduces the uncertainty in our results but also
enables us to include higher-order QNMs in the template, thereby mitigating the
charge-spin degeneracy. In this study, we generate merger-ringdown waveforms
based on the Kerr BH model and inject them into noise simulated from ET's
designed power spectral density (PSD). Subsequent analyses indicate that the
constraint on BH charge could be tightened to \(Q < 0.2\) at the \(90\%\) CL.

The remainder of this paper is organized as follows. In Sec.~\ref{sec:Method},
we introduce the Bayesian framework for constraining BH charge.
Section~\ref{sec:Results of GWTC-3} presents our findings derived from GWTC-3
data, confirming the charge-spin degeneracy of the remnant BH. In
Sec.~\ref{sec:Results of ET simulation}, we discuss results based on simulated
ET ringdown data, specifically addressing how the inclusion of higher multipole
modes can alleviate charge-spin degeneracy and yield more stringent constraints
on BH charge. Conclusions and discussions are provided in
Sec.~\ref{sec:Conslusions}.

\section{\label{sec:Method}Method}

In this section, we outline the waveform model and the GW data employed for
Bayesian inference.

During the ringdown phase, the GW signal emanates from the oscillations of the
remnant BH. In the context of a Kerr BH, the metric tensor field can be
decomposed into a background field and a perturbation term. This perturbation
term is a superposition of a set of eigenfunctions, leading to the solutions
known as QNMs \citep{Teukolsky1973,Berti2009}. Each QNM is characterized by a
damped sinusoid with a complex frequency,
\begin{equation}
\tilde{\omega}_{\ell m n} \equiv \omega_{\ell m n} - i \gamma_{\ell m n}, \quad
\gamma_{\ell m n} = \frac{1}{\tau_{\ell m n}},
\end{equation}
where \( \omega_{\ell m n} \) represents the oscillating frequency, \(
\gamma_{\ell m n} \) denotes the damping frequency, and \( \tau_{\ell m n} \) is
the damping time. Each QNM is distinguished by three indices, including \( \ell \) and \(
m \), which arise from the angular eigenfunctions, and \( n \), referred to as
the overtone number {that is associated with the damping time}. Generally, higher
overtone modes exhibit more rapid damping \cite{ringdown}.

In the case of a KN BH, electromagnetic perturbations come into play,
intertwining with gravitational perturbations. The specific calculations for
this are intricate and fall beyond the scope of this paper. As posited by the
no-hair theorem \cite{nohair}, the QNM frequencies are fully determined by the
mass \( M \), dimensionless spin \( a \), and charge-to-mass ratio \( Q \) of
the remnant BH, expressed as
\begin{equation}\label{eq:w}
\tilde{\omega}_{\ell m n} = \tilde{\omega}_{\ell m n}(M, a, Q).
\end{equation}
Here, we adopt the convention where \( a \) represents the dimensionless spin
and \( Q \) signifies the charge-to-mass ratio \( q/M \). Without losing
generality, we assume that $a$ and $Q$ are non-negative. Both \( a \) and \( Q
\) are constrained to lie between \( 0 \) and \( 1 \), subject to a GR-based
constraint
\begin{equation}\label{eq:aq}
a^2 + Q^2 \leq 1.
\end{equation}
{For the precise form of Eq.~(\ref{eq:w}), we rely on the numerical solutions
provided by \citet{KNqnm2,KNqnm3}. To elaborate further,   \citet{q} used an analytical fit for the numerical solutions \citep{KNqnm2,KNqnm3} and listed the fitting coefficients in their appendix. In this work we use their analytical fitting results}.

GW signal during the ringdown phase constitutes a superposition of various
possible QNMs. Typically, the \( \ell=2, m=2, n=0 \) mode (hereafter referred to
as the 220 mode) serves as the dominant contributor to the ringdown signal
\citep{Abbott2016b, 150914qnm, pyring1}. The 221 mode, characterized by a
similar oscillating frequency but a shorter damping time, has also been
considered in some previous QNM fittings for Kerr BHs   \citep{150914qnm2,150914qnm}.
Additionally, \citet{330} reported evidence supporting the existence of the 330
mode,  while analyses of \citet{210} support the existence of 210 and 320 modes when including precessing degrees of freedom. 
The disagreements are mainly between the remnant mass and spin of these two different ringdown analyses.
Since \citet{KNqnm2,KNqnm3} only provided the numerical solutions for the 220, 221 and 330 modes, we remain considering 220, 221 and 330 modes in our KN analysis. 
The higher modes, despite their theoretical underpinning
\cite{tianqin}, remain elusive in current LVK observations due to limited
detector sensitivity. Consequently, the template we employ for QNM fitting is,
\begin{equation}
\begin{aligned}
\label{eq:template}
h^{+}(t) - i h^{\times}(t) =& ~ A_{220} e^{-i (\tilde{\omega}_{220} (t-t_0) +
\varphi_{220})} {}_{-2} S_{220}(\iota, \varphi) \\
& + A_{221} e^{-i (\tilde{\omega}_{221} (t-t_0) + \varphi_{221})} {}_{-2} S_{221}(\iota, \varphi) \\
& + A_{330} e^{-i (\tilde{\omega}_{330} (t-t_0) + \varphi_{330})} {}_{-2} S_{330}(\iota, \varphi),
\end{aligned}
\end{equation}
where \( {}_{-2} S_{\ell m n} \) represents the spheroidal harmonics
\cite{Berti2006}, which depend on the spin polar angle \( \iota \) and the spin
azimuthal angle \( \varphi \), and \( t_0\) is the reference start time for the postmerger. {The real amplitude \( A_{\ell m n} \) and phase \(
\varphi_{\ell m n} \) of each mode are treated as free parameters. In Eq.~(\ref{eq:template}) we do not consider the contribution of the counter-rotating modes with the negative-\( m \) (see Refs.~\citep{q,counter-rotating1,counter-rotating2,counter-rotating3}). In addition we assume a non-precessing symmetry.} Moreover, the {polarization angle $\psi$ and the}
orientation of the GW event must also be specified to convert the observed GW
data into \( h^{+} \) and \( h^{\times} \) components \cite{ringdown}. In
summary, the parameters utilized for QNM fitting include,
\begin{equation}\label{eq:para}
\big\{ M, a, Q, A_{\ell m n}, \varphi_{\ell m n}, \iota, \psi \big\}.
\end{equation}
In this formulation, the spin azimuthal angle \( \varphi \) is subsumed into the
phase \( \varphi_{\ell m n} \). {Also, for our work based on the GWTC-3
data, we adopt the values of the sky location and start time reported in
Ref.~\citep{GWTC3TGR-release} for these events, so that they are not included in
Eq.~(\ref{eq:para}).} {Specifically, following \citet{GWTC3test}, we use a
reference time $t_0$ computed from an estimate of the peak of the strain,
$\sqrt{h^2_++h^2_{\times}}$, from the full IMR analyses.}

The prevailing model for GW data analysis is based on Kerr-like BHs, premised on
the expectation that BHs are electromagnetically neutral. Nonetheless, even when
focusing solely on ringdown data, introducing charge as an additional parameter
yields results comparable to those from the Kerr BH model. This arises from the
strong degeneracy between the spin and charge of the remnant BH, allowing the
effects of spin to be partially offset by an appropriate charge component (as
elaborated in Sec.~\ref{sec:Results of GWTC-3}). Consequently, spin estimates from prior analyses
employing the Kerr BH model can be interpreted as a form of ``effective spin''
if there is a nonvanishing charge. Employing a KN BH model for data fitting
allows us to derive charge distributions for the remnant, thereby establishing
upper limits on its charge. {The constraints
obtained are model-independent and serve to rule out scenarios involving highly
charged remnants, although the emergence of charge in these KN
fits is most likely an artifact of charge-spin degeneracy.}

Building upon the GWTC-2 dataset, which includes credible GW events up to the
O3a observing run of the LVK Collaboration, \citet{q} examined the potential
influence of charge in the merger-ringdown signals. Their analysis revealed the
presence of charge-spin degeneracy and also included a ``null test." In this
test, they assumed the validity of the Kerr hypothesis and incorporated a Kerr
BH prior for the final spin and mass, based on previous IMR results. Utilizing
the derived posterior distributions of charge,  an upper limit of \( Q < 0.33 \)
for GW150914 at the \( 90\% \) CL was obtained.

In the present study, we focus on five events cataloged in GWTC-3: GW191109,
GW191222, GW200129, GW200224, and GW200311. These events have been specifically
discussed in Ref.~\citep{GWTC3test} due to their credible ringdown signals. The
parameters of their remnant BHs are well-constrained relative to the prior, and
the Bayesian evidence supports the existence of a signal over mere Gaussian
noise. Furthermore, the sky-frame orientations of these events have been
accurately determined. For these five events, the LVK
Collaboration~\citep{GWTC3test} provided the outcomes of Kerr BH analyses, which
serve as the foundation for our subsequent analyses.

Our analysis of KN BHs employs the \textit{pyRing} software package, a
time-domain Bayesian inference tool designed for analyzing ringdown signals
\citep{pyRing,Abbott2016b,150914qnm,pyring1}. {The sampling method that we
utilized is the \textit{CPNest} algorithm \cite{cpnest}, which estimates the
evidence and obtains the posterior distribution iteratively.} For our KN BH
analysis, we initially set uniform priors for \(M\), \(A_{\ell m n}\),
$\cos\iota$, \(\varphi_{\ell m n}\), and \(\psi\) as indicated in
Eq.~(\ref{eq:para}). Specifically, \(M\) is chosen in the range of
\([0,200]\,M_{\odot}\), \(A_{\ell m n}\) is chosen in the range of \([0,10]
\times 10^{-21}\), {\(\cos\iota\) is chosen in the range of \([-1,1]\)},
and both \(\varphi_{\ell m n}\) and \(\psi\) in the range \([0,2\pi]\). The
parameters \(a\) and \(Q\) are uniformly distributed within the quarter disc
defined by \(a^2 + Q^2 \leq 1\), \(a \geq 0\), and \(Q \geq 0\).
{Following Ref.~\citep{GWTC3test}, we assume fixed values for the sky
location and the start time.}

The posterior distribution of these parameters is then computed using Bayes'
theorem
\begin{equation}
p(\vec{\theta} \mid d, \mathcal{H}) = \frac{p(\vec{\theta} \mid \mathcal{H})
\cdot p(d \mid \vec{\theta}, \mathcal{H})}{p(d \mid \mathcal{H})},
\end{equation}
where \(\vec{\theta}\) represents the model parameters, \(d\) is the observed
data, and \(\mathcal{H}\) denotes the model. Here, \(p(d \mid \mathcal{H})\) is
the evidence, calculated iteratively via \textit{CPNest} \cite{nest}. This
evidence is instrumental for model comparison, expressed through the Bayes
factor when comparing model 1 and model 2,
\begin{equation}\label{eq:bayes}
\mathcal{B}_{1}^{2} = \frac{p(d \mid \mathcal{H}_{2})}{p(d \mid
\mathcal{H}_{1})} = \frac{\int p(d \mid \vec{\theta_{2}}, \mathcal{H}_{2})
p(\vec{\theta_{2}} \mid \mathcal{H}_{2}) \mathrm{d} \vec{\theta_{2}}}{\int p(d
\mid \vec{\theta_{1}}, \mathcal{H}_{1}) p(\vec{\theta_{1}} \mid \mathcal{H}_{1})
\mathrm{d} \vec{\theta_{1}}}.
\end{equation}

For the GWTC-3 ringdown data, we incorporate both the 220 and 221 modes into our
model. Given that the 330 mode is considerably weaker and usually falls below
the noise level of LVK detectors, it is excluded from our KN analysis. {In the Kerr model analysis, \citet{GWTC3test} did not find preference for the 330 mode in the five GWTC-3 events that are used in this work.}
Our analysis employs ringdown data sampled at a rate of \(16\)~kHz, which
surpasses the commonly used \(4\) kHz rate. We utilize \(0.4\) s of data
following the merger for autocorrelation function (ACF) estimation to extract
the signal, and the full \(32\) s of data to estimate the PSD.
{To estimate the SNR, $\rho$, of a signal $h(t)$, we have
\begin{equation}\label{eq:snr}
\rho^2=h(t)\mathcal{C}^{-1}h(t)^{\intercal},
\end{equation}
where $\mathcal{C}$ is the auto-covariance matrix, which is the Toeplitz form of
the ACF.}

\section{\label{sec:Results of GWTC-3}Results from GWTC-3}

In this section, we delineate the outcomes of our KN BH analysis for the five GW
events selected from GWTC-3. A central aspect of our discussion will revolve
around the phenomenon of charge-spin degeneracy.

\begin{figure}
\centering
  \includegraphics[height=8cm]{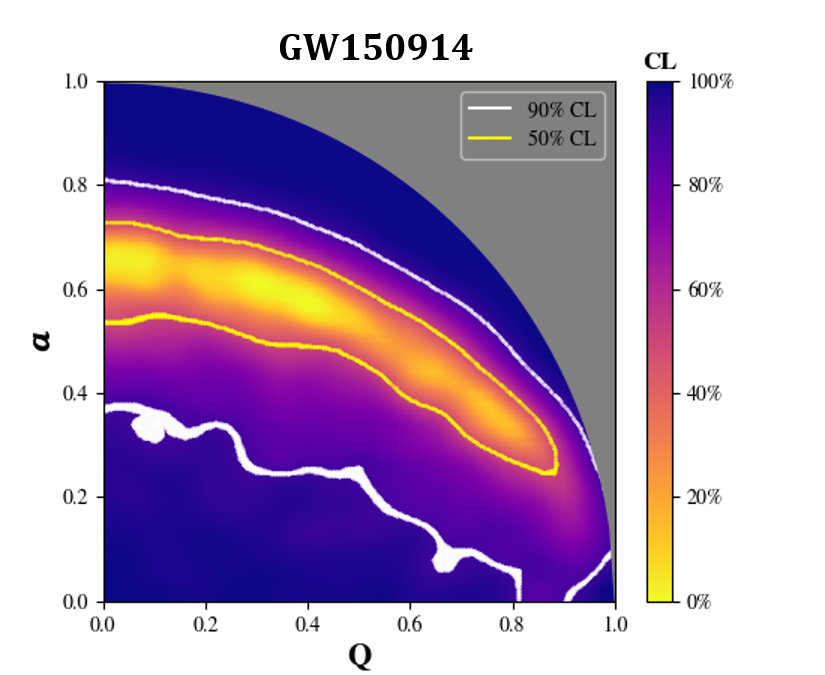}
\caption{\label{fig:fig2} Charge-spin posterior distribution for GW150914.
Different colors signify varying CLs within the distribution. The \(90\%\) CR
and \(50\%\) CR are demarcated by white and yellow lines, respectively.
The gray region violates the constraint \(a^2 + Q^2 \leq 1\). Note that the mass
parameter exhibits minimal correlation with the charge and is thus not
displayed.
}
\end{figure}

Differently from the approach taken by \citet{q}, our analysis employs a
sampling rate of \(16\) kHz. Figure~\ref{fig:fig2} displays the charge-spin
distribution we obtained for the remnant BH of GW150914. {The \(90\%\)
credible region (CR)} in our findings closely aligns with early work~\citep{q},
{while our result shows a slightly narrower distribution mainly due to the
16\,kHz sampling rate.} Employing the same methodology, we establish an upper
limit on the remnant BH charge of \(Q < 0.35\) at \(90\%\) CL. Further, in the
\(Q \rightarrow 0\) limit, the final spin approximates \(0.67\), aligning with
IMR results based on the Kerr BH model. The final mass, although not depicted in
the figure, also concurs with these Kerr-based IMR outcomes. Subsequent analyses
are performed on the selected events from GWTC-3.

\begin{figure*}[ht]
\begin{minipage}{0.5\textwidth}
\centering
  \includegraphics[height=7.5cm]{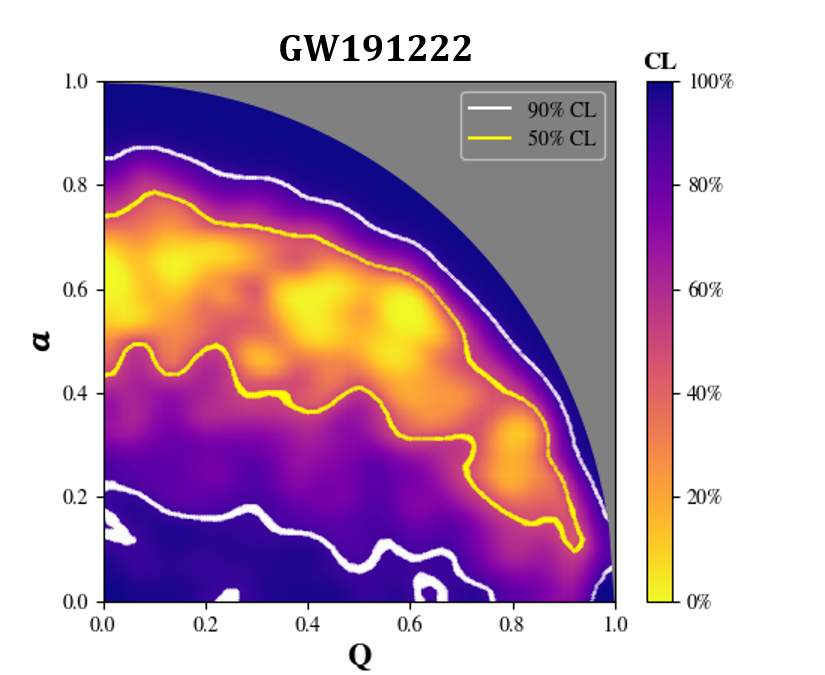}
\end{minipage}\hfill
\begin{minipage}{0.5\textwidth}
\centering
  \includegraphics[height=7.5cm]{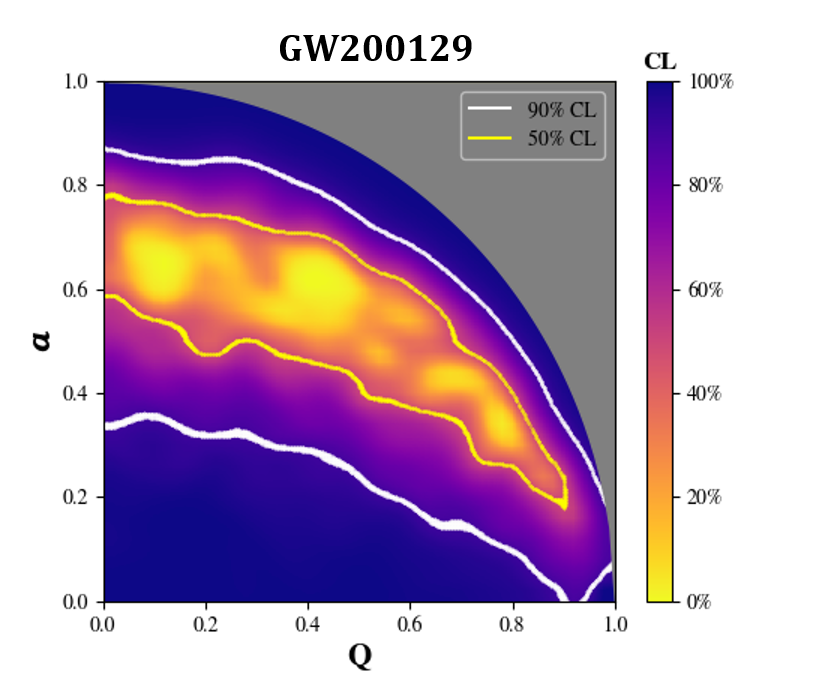}
\end{minipage}\hfill
\begin{minipage}{0.5\textwidth}
\centering
  \includegraphics[height=7.5cm]{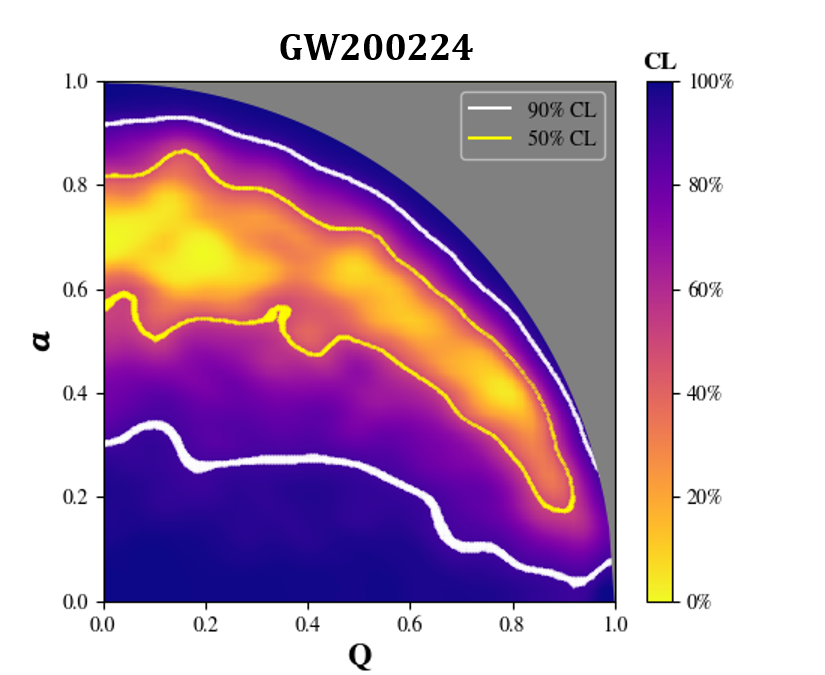}
\end{minipage}\hfill
\begin{minipage}{0.5\textwidth}
\centering
  \includegraphics[height=7.5cm]{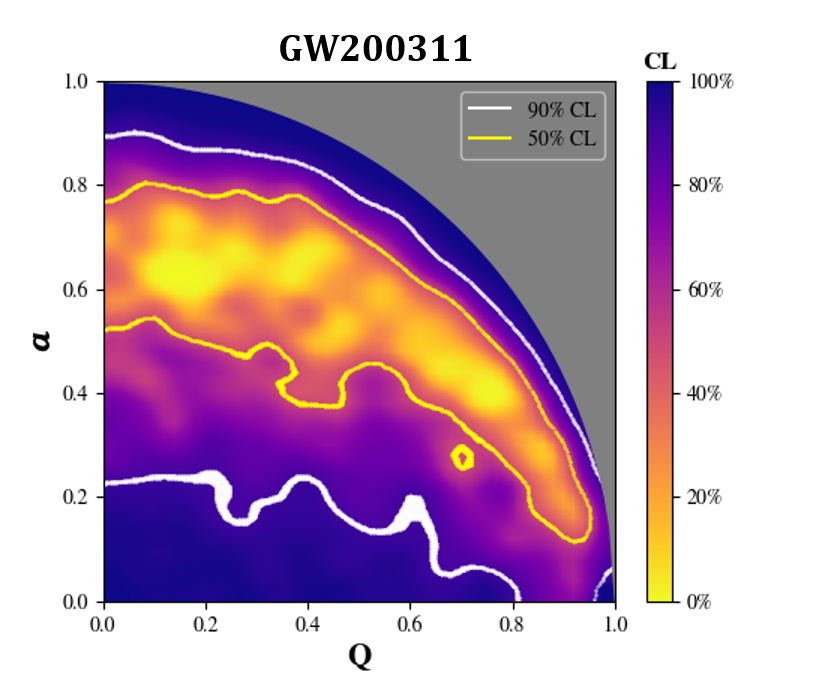}
\end{minipage}\hfill
\begin{minipage}{1\textwidth}
\centering
  \includegraphics[height=7.5cm]{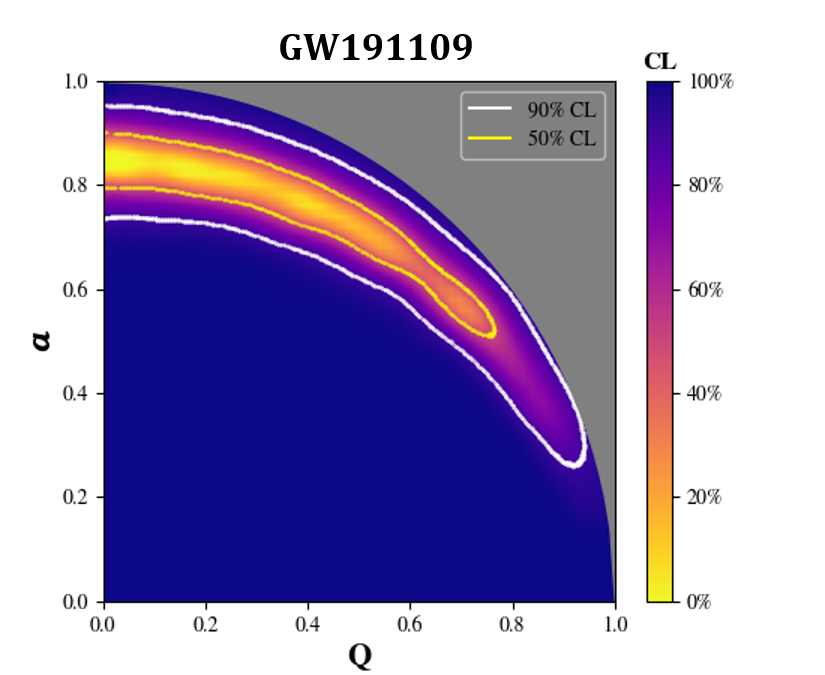}
\end{minipage}
\caption{\label{fig:fig3} Same as Fig.~\ref{fig:fig2}, but for GW191222,
GW200129, GW200224, GW200311 and GW191109.}.
\end{figure*}

\begin{figure*}[ht]
  \begin{minipage}{1\textwidth}
  \centering
    \includegraphics[height=7.5cm]{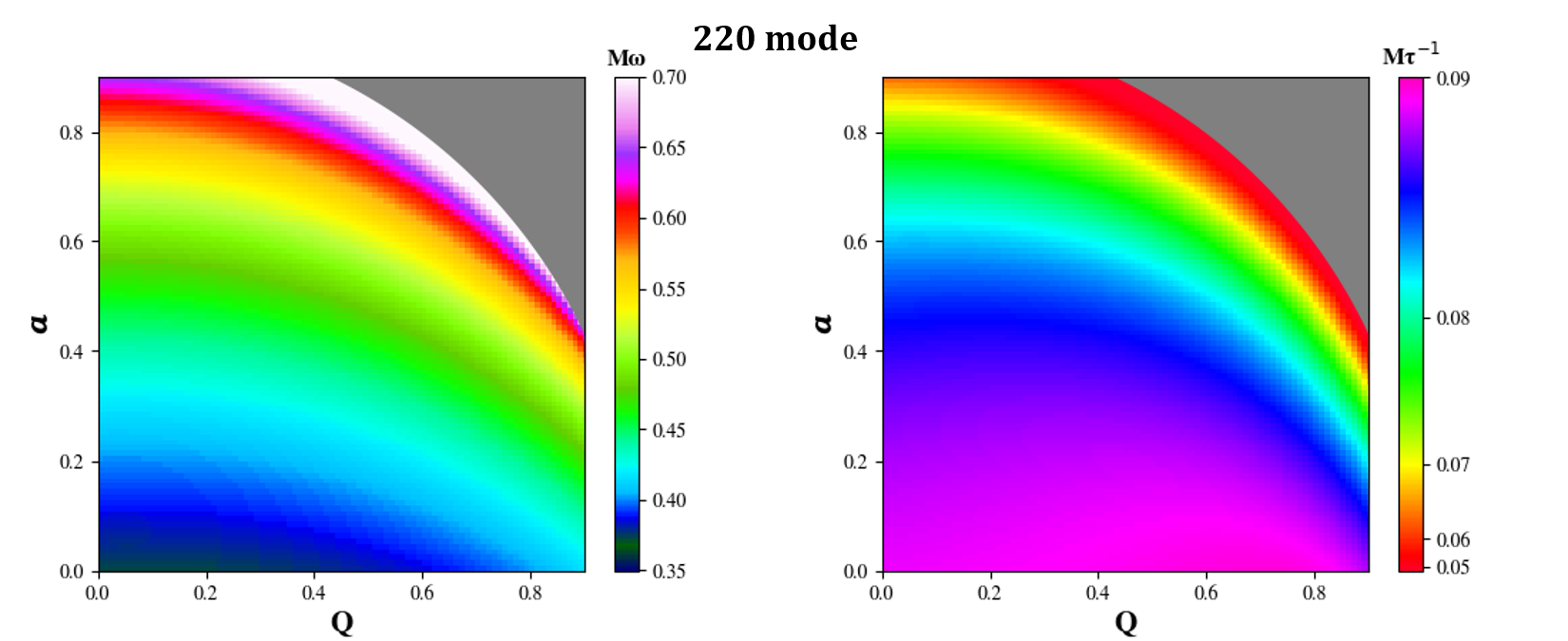}
  \end{minipage}\hfill
  \begin{minipage}{1\textwidth}
  \centering
    \includegraphics[height=7.5cm]{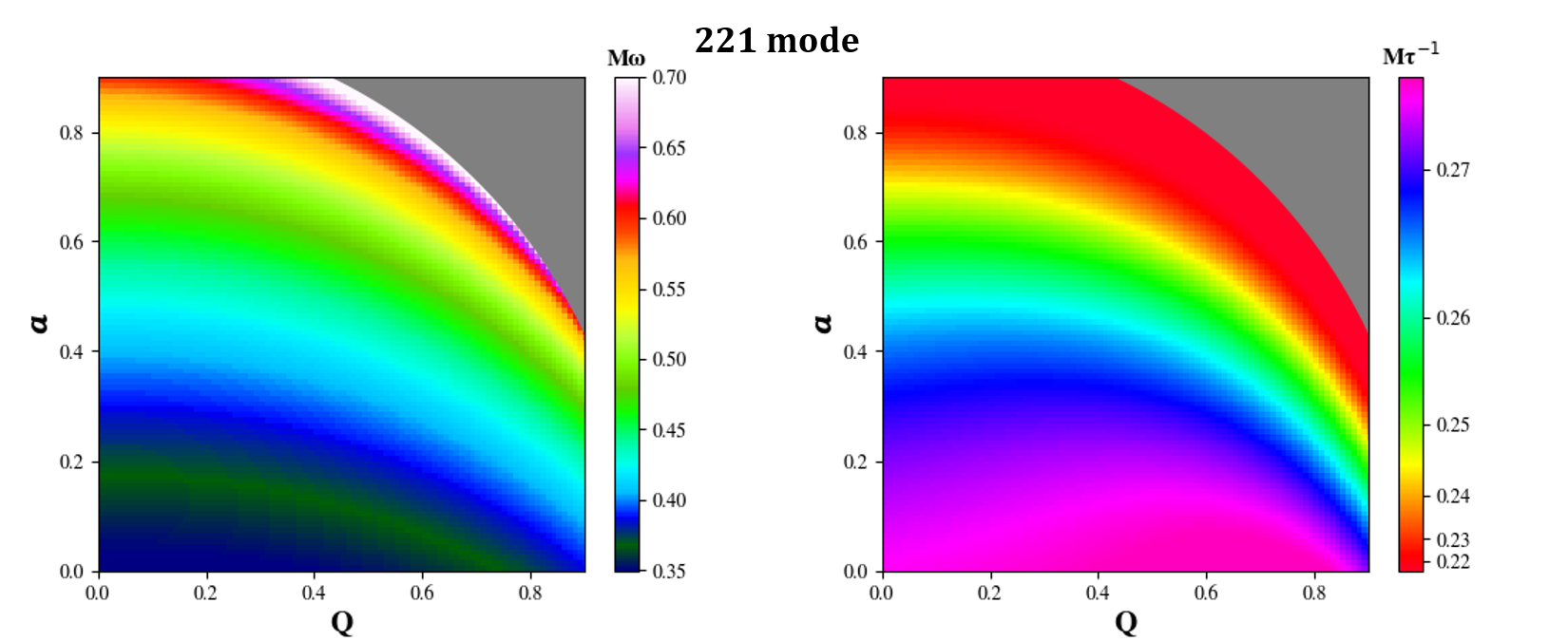}
  \end{minipage}\hfill
  \begin{minipage}{1\textwidth}
  \centering
    \includegraphics[height=7.5cm]{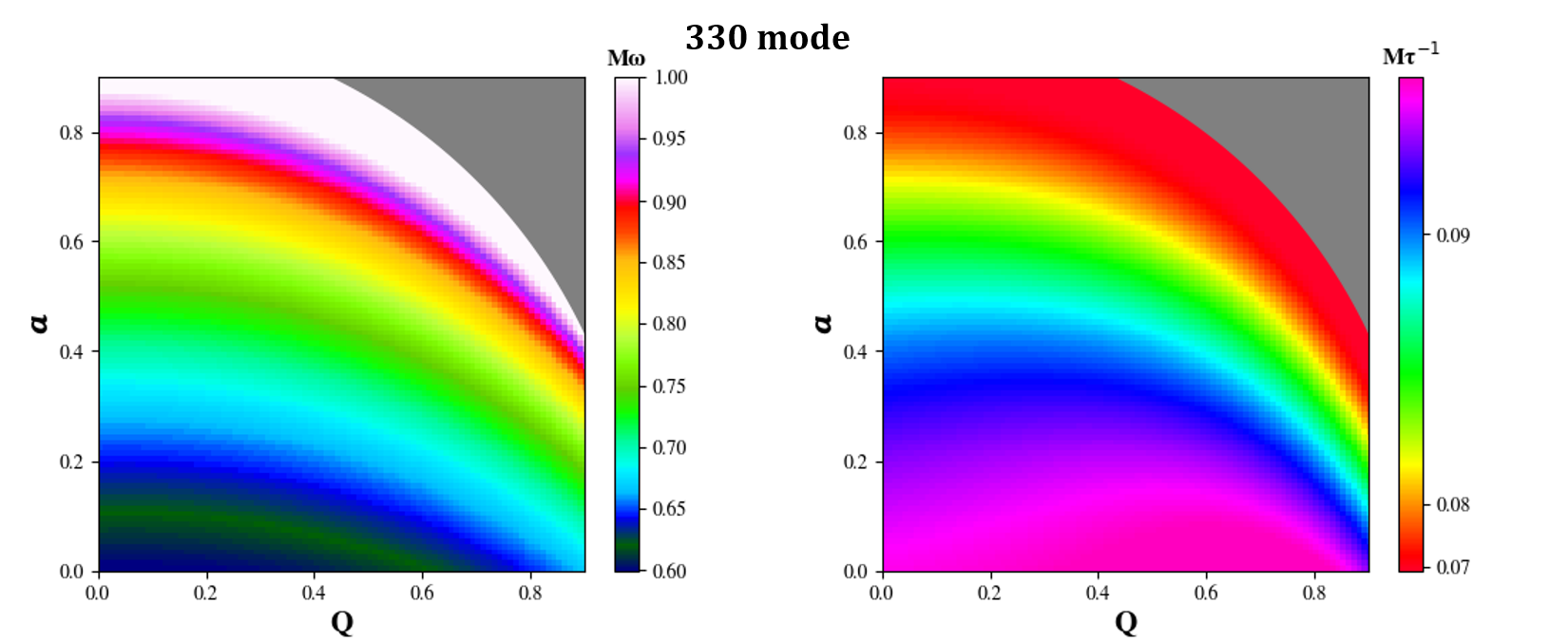}
  \end{minipage}
  \caption{\label{fig:fig4} QNM frequencies for the 220, 221, and 330 modes. The
  left panels display the oscillating frequencies, while the right panels show the
  damping frequencies. Different colors signify different frequency values. The
  gray region violates the constraint \(a^2 + Q^2 \leq 1\).}
  \end{figure*}

\begin{table*}[ht]
  \caption{\label{tab:table2} Constraints on BH charges, denoted by
  \(Q_{\text{max}}\), for five GWTC-3 events and two GWTC-2 events, given at the
  \(90\%\) CL for both Gaussian IMR and flat bounded IMR priors. Also listed are
  the logarithmic Bayes factors between the KN BH and Kerr BH models, denoted by
  \(\ln\mathcal{B}_{\text{Kerr}}^{\text{KN}}\), for both the uniform-prior and IMR
  prior cases (null tests), as well as the SNRs for the ringdown data {(calculated also from \textit{pyRing})}.}
  \begin{ruledtabular}
  \renewcommand\arraystretch{1.35}
  \begin{tabular}{c c c c c c }
  Event   & $Q_{\rm max}$ (Gaussian prior) & $Q_{\rm max}$(flat bounded prior) &
  $\ln\mathcal{B}_{\rm Kerr}^{\rm KN}$ & $\ln\mathcal{B}_{\rm Kerr}^{\rm
  KN}$(null) &  Ringdown SNR \\\hline
  GW191109 & $0.77$            & -   & $-1.4$ & $-0.5$                     & $12.6$         \\ 
  GW191222 & $0.50$            & $0.59$   & $0.3$ & $-0.4$                  & $6.4$          \\ 
  GW200129 & $0.37$            & $0.45$ & $0.3$ & $-0.6$                    & $13.0$         \\ 
  GW200224 & $0.37$            & $0.53$  & $-1.1$ & $-0.7$                   & $10.7$         \\ 
  GW200311 & $0.47$            & $0.57$   & $-0.8$ & $-0.5$                  & $7.9$          \\ 
  GW150914 & $0.37$            & $0.35$   & $-0.6$ & $-0.7$                  & $12.6$         \\ 
  GW190521\_07 & $0.40$            & $0.41$  & $-0.2$ & $-0.8$                  & $9.6$         \\ 
  \end{tabular}
  \end{ruledtabular}
  \end{table*}

Figure~\ref{fig:fig3} presents the charge-spin distributions of the remnant BHs
for the five events featured in GWTC-3. Among them, GW200129 and GW200224 yield
more stringent constraints on both charge and spin, while GW191222 and GW200311
offer weaker constraints due to their relatively low ringdown SNRs.
Intriguingly, the most stringent yet puzzling constraints emanate from GW191109.
Although there exists a strong correlation between charge and spin for this
event, the constraint on charge is surprisingly lax. No discernible improvement
in SNR accounts for this anomaly (see Table~\ref{tab:table2}). The ringdown SNR
for GW191109 is \(12.6\), comparable to that of GW150914. We will delve into
this particular case in further detail later. 

To assess the validity of our KN BH model, we concurrently perform analyses
under the assumption of the Kerr hypothesis, setting the charge to zero.
Subsequently, we compute the Bayes factor between the KN and Kerr BH models as
outlined in Eq.~(\ref{eq:bayes}). The results indicate a slight preference for
the Kerr BH model, as most of the log Bayes factors are negative (see
Table~\ref{tab:table2}). This outcome aligns with similar findings from GWTC-2
events \cite{q}, lending support to the prevailing Kerr hypothesis that remnant
BHs are uncharged.

The charge-spin distribution for GW191109 exhibits a ``belt-like" shape, a stark
contrast to the more diffuse distributions typically observed. This unique
formation is a direct consequence of the charge-spin degeneracy, wherein various
points in the charge-spin parameter space yield indistinguishable QNM
frequencies, thus forming a belt-like distribution in the results. {Although} this phenomenon is also manifest in the other four events, it remains less
conspicuous due to the larger associated uncertainties. 

Notably, our ringdown analysis for GW191109 reveals a discrepancy when compared
to previous IMR results in the Kerr BH model. The IMR result for GW191109 gives
\(M_f = 132.7^{+21.9}_{-13.8} M_{\odot}\) and \(a_f = 0.60^{+0.22}_{-0.19}\).
However, in the \(Q \rightarrow 0\) limit, our analysis yields \(M_f =
184^{+12}_{-12} M_{\odot}\) and \(a_f = 0.85^{+0.05}_{-0.06}\). These findings
align with the ringdown analysis by the LVK Collaboration~\citep{GWTC3test},
where the Kerr BH model was employed to fit the ringdown signal, resulting in
\(M_f = 179.0^{+23.7}_{-21.7}\) and \(a_f = 0.81^{+0.08}_{-0.14}\)  at \(90\%\)
CL. {This discordance between the ringdown and inspiral signals points to an
inconsistency, an issue also noted in Ref.~\citep{GWTC3test},  where they attributed it to the non-Gaussian
noise.} For the remaining four events, our ringdown-derived results are
consistent with earlier IMR analyses.

We now delve into the charge-spin degeneracy alluded to earlier, extending the
preliminary discussion  by \citet{q}. Figure~\ref{fig:fig4} vividly illustrates
how the QNM frequencies vary with both the final spin \( a \) and charge \( Q
\), based on the analytical fit results from \citet{q}. Specifically, the dimensionless frequencies \(
M\omega \) and \( M\tau^{-1} \) serve as functions of the dimensionless spin \(
a \) and charge-to-mass ratio \( Q \). As depicted in Fig.~\ref{fig:fig4}, the
oscillating and damping frequencies of the 220 mode exhibit a strikingly similar
dependence on \( a \) and \( Q \), giving rise to the belt-like distribution in
the \( Q \mbox{-} a \) plane.

{Introducing additional QNMs} into the fitting—such as the 221 and 330
modes—can alleviate this degeneracy. As Fig.~\ref{fig:fig4} reveals, different
damping frequencies manifest distinct dependencies on \( a \) and \( Q \). The
inclusion of higher modes enriches the information available for \( a \) and \(
Q \) estimation, thereby yielding more precise constraints. However, due to the
limited sensitivity of current LVK observations, accurate extraction of these
higher modes remains challenging. In our analysis based on GWTC-3 data, we
confined ourselves to the 220 and 221 modes, resulting in pronounced degeneracy.
Next-generation GW detectors like the ET are expected to facilitate the
inclusion of higher modes in the models, thereby breaking the degeneracy more
effectively. {Additionally, it is worth noting that this degeneracy is intrinsic
for small \( Q \) values, as the frequencies have weak dependence on \( Q \) when \( Q \) is near zero \citep{KNqnm2,KNqnm3}}. This implies that BHs with minimal charge
may evade detection through this method, setting a lower bound on the charge
constraints.

\begin{figure}[ht]
  \includegraphics[height=7.3cm]{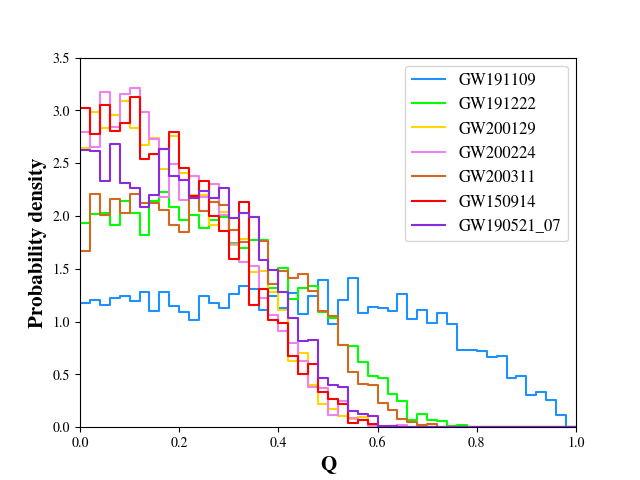}
\caption{\label{fig:fig5}Constraints on the final BH charge, according to the
Gaussian prior of spin from Kerr IMR analysis. Different lines represent the
posterior charge distributions of different events.}
\end{figure}

The charge and spin are correlated in the above results. To obtain more
stringent constraints on the charge of the remnant BH, we can constrain the spin
first. Since our method is based on the Bayesian inference, we could apply a
more stringent prior on the spin. Since the electromagnetic observations
\citep{galactic1,galactic2} support the Kerr hypothesis, a tentative choice is
to use the final spin result from IMR analysis, which is based on the Kerr BH
model. This procedure, i.e. using the Kerr prior to obtain the posterior
distribution of charge, is called the ``null test" in \citet{q}, as this prior
is based on the hypothesis that the final BH does not possess charge. Among the
obtained posterior distributions of charge, all the events except GW191109 show
a decreasing probability density from $Q=0$, signifying that they are very
likely to have negligible charge. GW191109, however, has a very broad charge
distribution. This is due to the strong correlation of charge and spin, where
they cannot be well constrained at the same time. These results are shown in
Fig.~\ref{fig:fig5}. The results of GW150914 and GW190521\_07 (i.e.
GW190521\_074359) are also presented for comparison. Here we choose the form of
the priors as Gaussian, instead of the flat bounded prior used by \citet{q}. We
chose a Gaussian prior because it matches the distribution of the parameters
from IMR analysis more accurately.  {However, from another point of view,
the Gaussian prior will introduce more information compared to a uniform
distribution.} Furthermore, the final spin distribution of GW191109 from Kerr
ringdown analysis falls outside the range of its IMR result, which makes it
unsuitable to adopt a flat bounded prior. We choose the value of the priors
according to the results of the LVK Collaboration \citep{GWTC3test}.

We use the $90\%$ CL limit in the posterior charge distributions as the upper
bound on charge, for the selected events separately. The upper limits on charge
are listed in Table~\ref{tab:table2}, where we also show the limits given by
flat bounded prior for comparison. For results of GWTC-3, GW200129 and GW200224
give the strongest constraint $Q<0.37$. This constraint is at the same level as
the GWTC-2 results from \citet{q}. GW191222 and GW200311 have lower ringdown
SNRs and can only give limits at about $Q<0.5$. GW191109 gives the worst
constraint in our selected GWTC-3 events. One may argue that this is due to the
discrepancy between the IMR Kerr prior and KN results for GW191109, i.e. we
choose a prior smaller than its Kerr spin from ringdown analysis, and the
charge-spin degeneracy causes the nonzero value of charge (the charge is fitted
to compensate for the spin), thus leading to this broad distribution. However,
as will be shown in the following, this is not related to the choice of prior.

\begin{figure*}
\begin{minipage}{0.5\textwidth}
\centering
  \includegraphics[height=7.5cm]{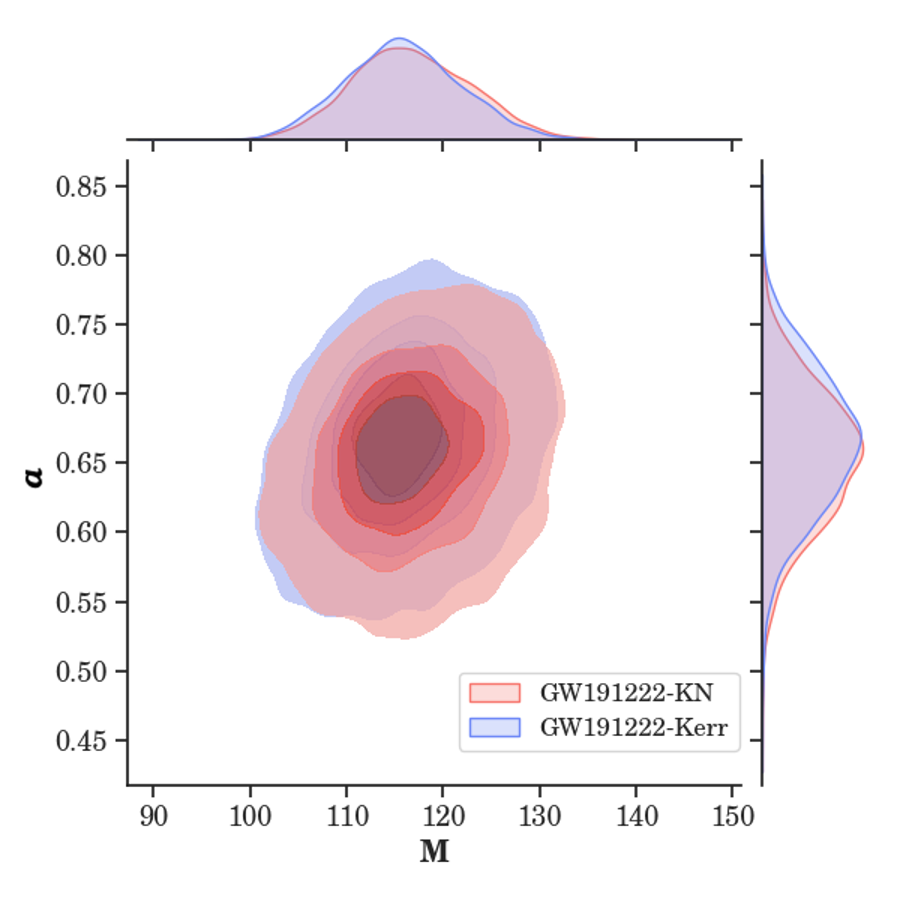}
\end{minipage}\hfill
\begin{minipage}{0.5\textwidth}
\centering
  \includegraphics[height=7.5cm]{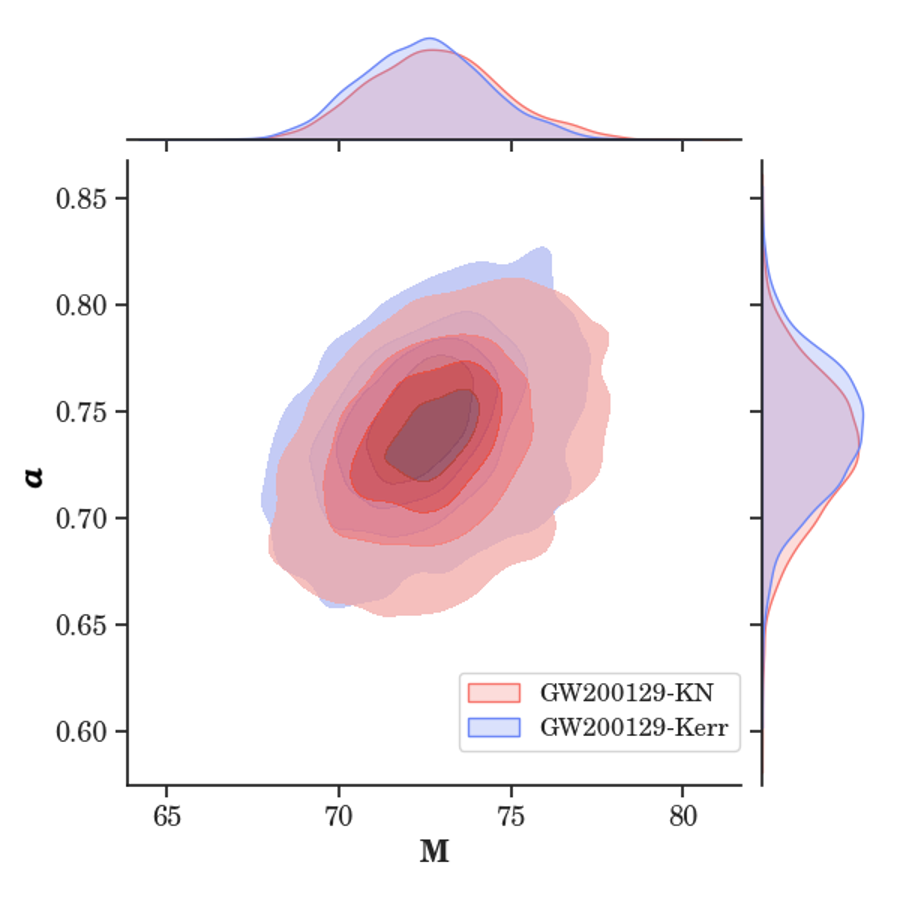}
\end{minipage}\hfill
\begin{minipage}{0.5\textwidth}
\centering
  \includegraphics[height=7.5cm]{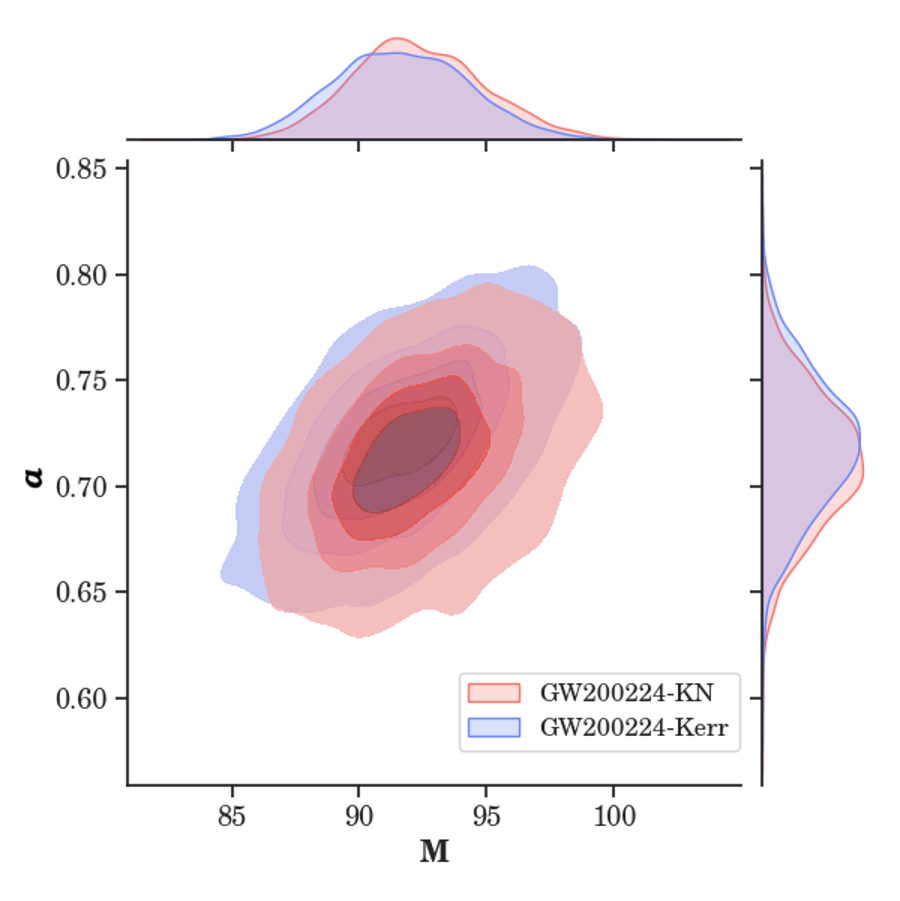}
\end{minipage}\hfill
\begin{minipage}{0.5\textwidth}
\centering
  \includegraphics[height=7.5cm]{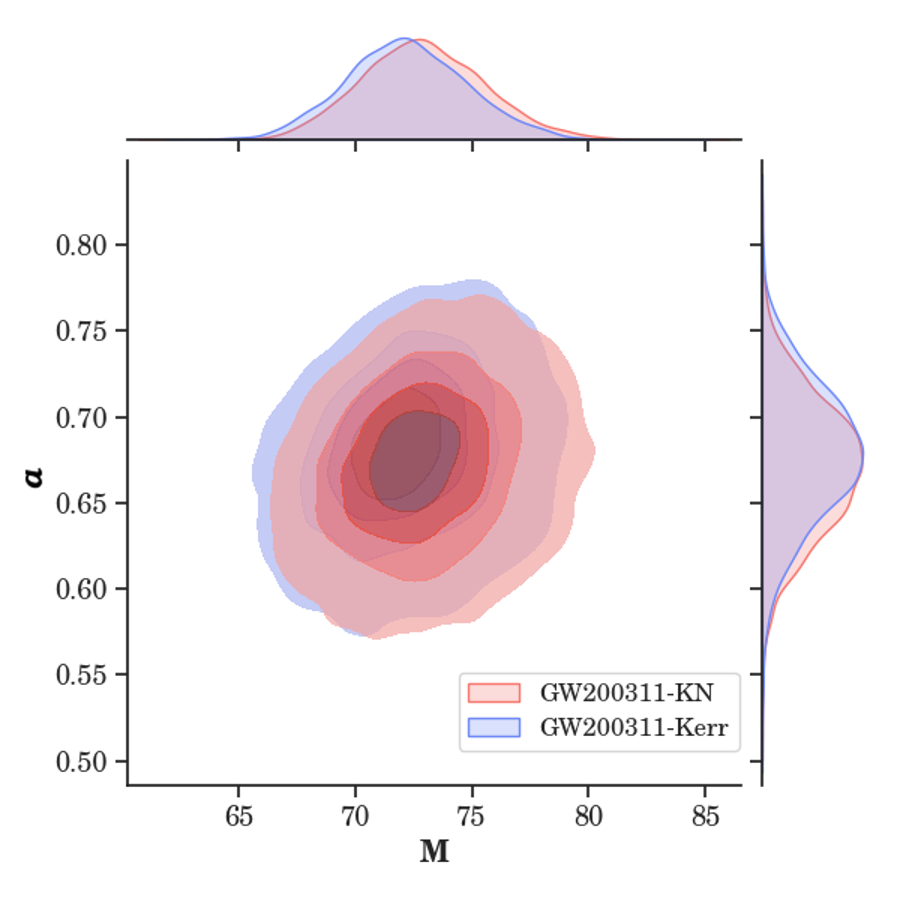}
\end{minipage}\hfill
\begin{minipage}{1\textwidth}
\centering
  \includegraphics[height=7.5cm]{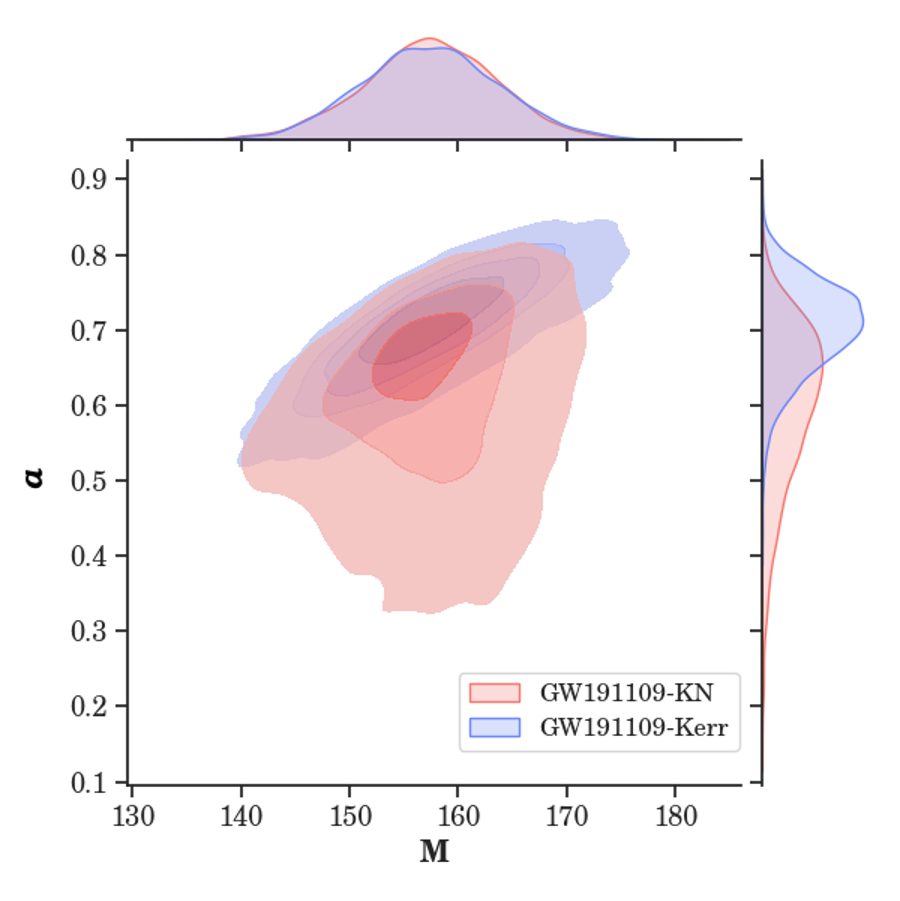}
\end{minipage}
\caption{\label{fig:fig6}Distributions of final mass \( M \) (in the unit of \(
M_{\odot} \)) and final spin \( a \) for five GWTC-3 events, under the Kerr IMR
prior. The blue and red shadows represent the probability density distributions
in the Kerr BH model and in the KN BH model, respectively. The top and right
panels display the projected distributions of final mass and final spin.}
\end{figure*}

{Constraints here are based on the IMR-based prior and KN BH model. We
choose this prior according to the Kerr BH hypothesis.} To assess the validity
of this assumption, we perform a companion analysis using the Kerr BH model and
compare the results with those obtained using the KN BH model. In both
situations, we adopt the Kerr prior {(i.e.\ IMR-based prior)}.
Figure~\ref{fig:fig6} presents the posterior distributions of the remnant mass
and spin for the five GW events under consideration. For all events, except
GW191109, the obtained remnant mass and spin are visually identical for the two
models.  {The spin posterior is larger in the Kerr case because of the
degeneracy between the final mass and final spin. The spin distribution in
the Kerr BH model may be slightly higher than that in the KN BH model},
which can be attributed to the charge-spin degeneracy.  We also compute the
evidence for the Kerr BH model and calculate the Bayes factors for the null
test. {The log Bayes factors are all negative in our results, suggesting that the
Kerr model provides a slightly better fit to the data. However,
it should be noted that these values are quite small, indicating that a charged
BH cannot be ruled out.}  These results are also summarized in
Table.~\ref{tab:table2}.

For GW191109, the spin obtained from the Kerr BH model is notably higher than
the spin obtained from the KN BH model. In both cases, we adopt the IMR Kerr
prior, implying that the observed difference is not a result of prior selection
but rather originates from the intrinsic differences between the Kerr BH and KN
BH models.  On the other hand, the distribution of the spin is larger in the
Kerr case, which is due to the degeneracy between the final mass and final spin.
{This may simply be an effect of charge-spin degeneracy. Furthermore,
GW191109 is a unique event, whose results are very likely influenced by its
intrinsic noises.} The logarithmic Bayes factor for this event is \( -0.5 \),
suggesting that {the KN result is not strong.  Consequently, we are
inclined to attribute the discrepancy in the final spin between the Kerr BH and
KN BH models to the non-Gaussian noise, rather than the presence of charge.}

\section{\label{sec:Results of ET simulation}ET simulation}

The charge-spin degeneracy prevents us from giving stronger constraints on the
BH charge. To break the degeneracy, as illustrated in Sec.~\ref{sec:Results of
GWTC-3}, incorporating higher QNMs might be a solution. However, the requirement to detect
higher modes places higher demands on the SNR of the ringdown signal. Therefore,
in this section, we analyze the simulated GW ringdown data, using the noise of
ET. ET is one of the next-generation GW observatories planned for construction
in the 2030s. Designed as an equilateral triangle with arms extending $10$ km,
ET aims for a tenfold enhancement relative to LVK detectors in sensitivity,
thereby paving the way for more precise QNM analyses \cite{ETtriangle}. 

To begin, we simulate the ringdown waveform using parameters that mirror those
of the GW150914 event, with $M_f = 68.0 \, M_\odot$, $a_f = 0.67$, and the
polarization angle $\psi = 0.0$.  We incorporate three QNMs---220, 221, and
330---with amplitude levels consistent with the GW150914 signal. In the Kerr BH
framework, we set the amplitude ratios \(A_{220}=2.0\), \(A_{221}=3.0\), and
\(A_{330}=0.2\), guided by theoretical values that relate the 22 (\(l=2, m=2\))
and 33 modes \cite{ratio}.\footnote{Although the relative amplitude for the 221
mode is not theoretically determined in Ref.~\citep{ratio}, we adopt our own
fitting results from GW150914. Specifically, the maximum likelihood set of
values yielded a ratio of \(A_{220}:A_{221} = 2:2.9\). This approximate ratio
was further corroborated by our GWTC-3 analysis for all events.} Subsequently,
we embed this simulated waveform into the Gaussian noise spectrum of ET. {The ET noise spectrum used in our study is provided by \textit{GW-Toolbox} \cite{Yi:2021wqf}.\footnote{\hyperlink{http://www.gw-universe.org/}{http://www.gw-universe.org/}}} The SNR
for our simulated event approximates $270$.

\begin{figure*}[ht]
  \begin{minipage}{0.5\textwidth}
  \centering
    \includegraphics[height=7.5cm]{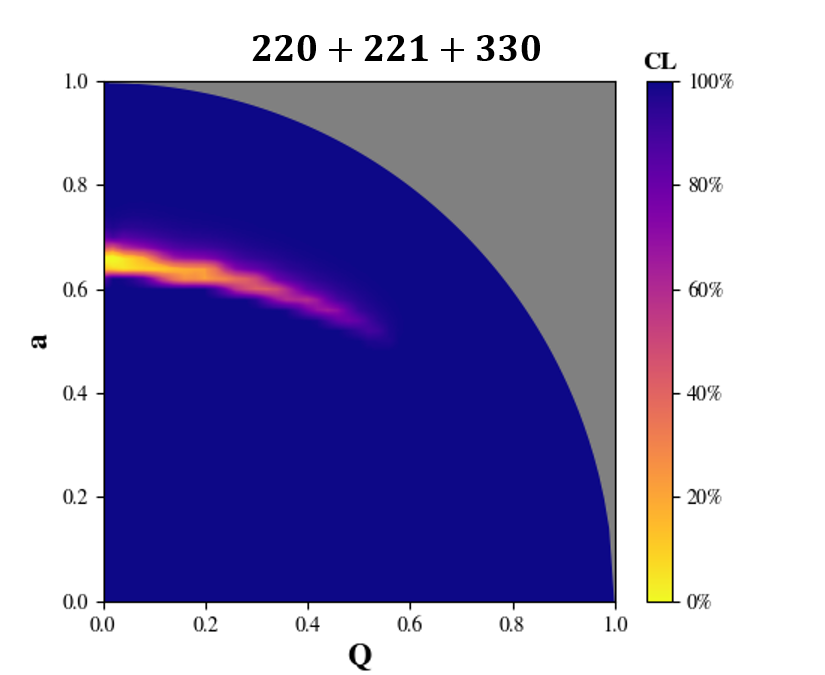}
  \end{minipage}\hfill
  \begin{minipage}{0.5\textwidth}
  \centering
    \includegraphics[height=7.5cm]{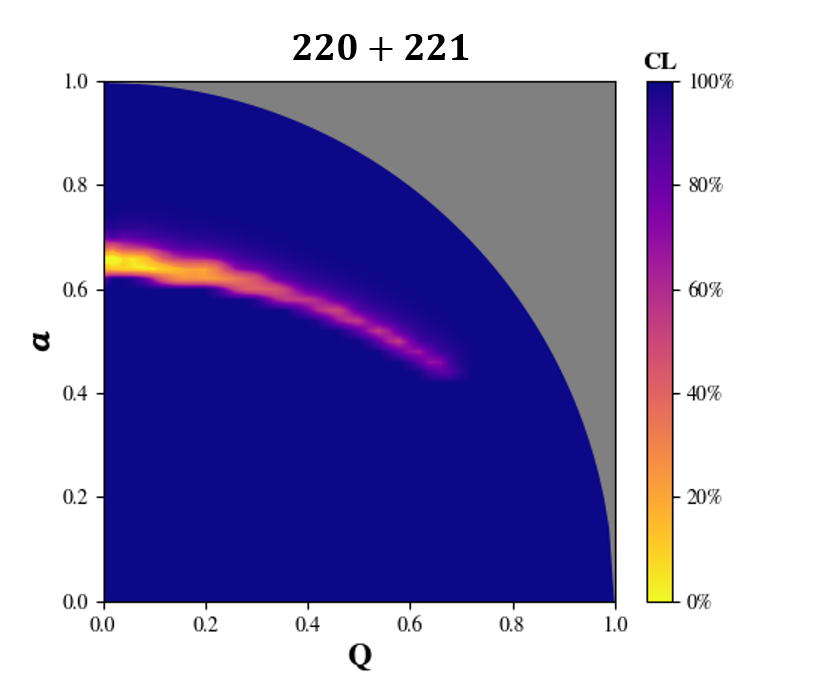}
  \end{minipage}\hfill
  \caption{\label{fig:fig7}Same as Fig.~\ref{fig:fig2}, but for the simulated ET
  ringdown data, where the injected final spin and the final charge of the remnant
  are $a=0.67$ and $Q=0$, respectively. {The left panel presents the posterior when incorporating the 220, 221 and 330 modes collectively. In comparison, the right panel is for only considering 220 and 221 modes.}
  }
  \end{figure*}

We proceed with Bayesian inference using the KN BH model on the simulated ET
ringdown data, employing a methodology consistent with our earlier approach. The
only modification is the inclusion of an additional higher mode, specifically
the 330 mode, in the fitting procedure. The resulting charge-spin distribution
is depicted in {the left panel} of Fig.~\ref{fig:fig7}. The posterior distribution for charge
predominantly converges toward zero. When juxtaposed with the GWTC-3 results,
the error margins are markedly reduced, and the correlation between \( Q \) and
\( a \) is more discernible. {However,} the belt-like distribution persists,
indicating that the charge-spin degeneracy is not entirely eliminated, the
degeneracy is noticeably lessened.

We further investigate the impact of incorporating higher modes into our
analysis. As delineated in Sec.~\ref{sec:Results of GWTC-3}, the inclusion of
higher modes can mitigate the charge-spin degeneracy. {We now focus on the extent of this effect in incorporating the 330 mode, apart from the improvement in SNR. The right panel of Fig.~\ref{fig:fig7} shows the outcome when considering only the 220 and 221 modes, under
the KN BH model for the simulated ringdown signal. Compared to the left panel, where the 220, 221, and 330 modes are all included in the fitting, adding the 330 mode exhibits slightly weaker degeneracy effects. } 
Although we anticipate that even more
precise results could be achieved by including still higher modes like the 440
mode, the requisite QNMs for the KN BH model remain to be computed.

\begin{table*}
  \caption{\label{tab:table4}Fitting results for simulated ringdown signals with
  varying remnant charges. The columns represent the injected remnant charge $Q$,
  the logarithm of the Bayes factor between the KN and Kerr BH models
  $\ln\mathcal{B}_{\rm Kerr}^{\rm KN}$, the prior for the final mass and spin
  utilized in the null test at the \(90\%\) CL,\footnote{In actual
  scenarios, the priors for mass and spin would be determined by IMR analysis.
  However, as we have not simulated the inspiral phase, we employ the Kerr BH
  fitting results for this ringdown signal as a surrogate prior. Though this
  approach lacks strict Bayesian rigor, it serves as a reasonable approximation
  for the \(Q=0.2\) and \(Q=0.4\) cases.}, the posterior distribution of charge
  at the \(1 \mbox{-} \sigma\) level in the null test and the SNRs for the simulated data.}
  \begin{ruledtabular}
  \renewcommand\arraystretch{1.35}
  \begin{tabular}{c c c c c c }
  Injected $Q$    & $\ln\mathcal{B}_{\rm Kerr}^{\rm KN}$    & Kerr mass
  ($M_\odot$)    & Kerr spin           & Posterior $Q$ (null) & {Ringdown SNR}  \\\hline
  0.0 & $-0.6$ & $68.7^{+0.5}_{-0.6}$ & $0.670^{+0.015}_{-0.015}$ &$0.09^{+0.05}_{-0.09}$  &  {$267.8$} \\ 
  0.2 & 0.2 & $68.3^{+0.7}_{-0.7}$   & $0.682^{+0.017}_{-0.019}$ &$0.17^{+0.10}_{-0.12}$  &  {$268.9$} \\ 
  0.4 & 0.7  & $67.3^{+0.5}_{-0.5}$  & $0.719^{+0.012}_{-0.012}$ & $0.24^{+0.13}_{-0.09}$  &  {$274.7$} \\ 
  \end{tabular}
  \end{ruledtabular}
  \end{table*}

We proceed to replicate the foregoing analysis using the Kerr BH model and also
compute the Bayes factor \(\mathcal{B}_{\text{Kerr}}^{\text{KN}}\). The
logarithm of the Bayes factor amounts to \( -0.6 \), signaling a preference for
the Kerr BH model over the KN BH model. Given that the signal was simulated
under the assumption of the Kerr BH model, this outcome aligns with our
expectation. However, the presence of charge-spin degeneracy raises the question
of whether the Kerr BH model would still yield a reasonable fit even if the
actual signal had included a nonzero charge. To scrutinize this, we inject
ringdown signals with charges \( Q = 0.2 \) and \( Q = 0.4 \). The log Bayes
factors{, although small,} are positive in both cases, validating the robustness of our analysis.
The negative log Bayes factor emerges only when the BH is not substantially
charged, underscoring the utility of the Bayes factor in evaluating the Kerr
hypothesis in future ET observations. {The differences between these Bayes factors would be more significant when incorporating higher QNMs, where
the above illustration could be more informative.} These findings are summarized in
Table.~\ref{tab:table4}.

\begin{figure}[ht]
  \includegraphics[height=8cm]{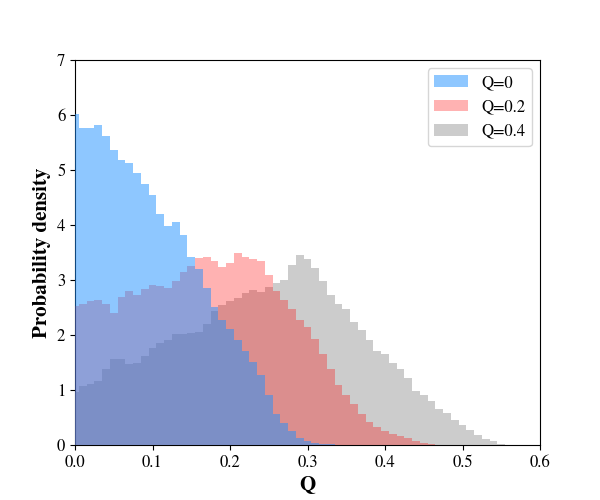}
\caption{\label{fig:fig9}Posterior distributions of the final BH charge for the
simulated ET ringdown data using the Kerr prior. The blue, red and gray shadows
represent the results with an injected remnant charge $Q=0$, $Q=0.2$ and
$Q=0.4$, respectively. }
\end{figure}

We next turn our attention to the so-called ``null test", in which we employ
Kerr-based priors to constrain the final spin and evaluate the upper limits on
the remnant charge. The priors used for the remnant mass and spin are outlined
in Table~\ref{tab:table4}, along with the derived posterior distributions for
the remnant charge. In addition to the Kerr BH scenario, we also examine cases
with injected charges \(Q=0.2\) and \(Q=0.4\) for comparative analysis. As
depicted in Fig.~\ref{fig:fig9}, these cases yield noticeably distinct charge
distributions, even when the priors are rooted in the Kerr hypothesis. This
outcome signals that the charge-spin degeneracy is only partially alleviated.
For the \(Q=0\) case, we constrain the remnant charge to \(Q<0.2\) at the
\(90\%\) CL. {If only considering the 220 and 221 modes, the constraint is about \(Q<0.3\) at
\(90\%\) CL, which is at the same level as the GWTC-3 outcomes}

Our constraint of \(Q<0.2\) on the remnant charge may appear somewhat modest,
especially when contrasted with existing LVK constraints,  \(Q \lesssim 0.3\),
despite the tenfold enhancement in ET's sensitivity. \citet{Cardoso2016}
theorized that more stringent constraints could be achieved for high-spin
events. Confirming this, our simulation with \(a=0.9\) yielded an upper limit of
\(Q<0.15\). However, this tighter constraint may in part be attributed to the
\(a^2+Q^2 \leq 1\) restriction. Indeed, our simulation results fall short of the
predictions made by \citet{Cardoso2016}, who posited an upper limit on \(Q\) of
less than \(0.1\). We attribute this discrepancy to the differing methodologies.
{Different from the employed Fisher information matrix approach with a
fixed spin in \citet{Cardoso2016}}, our Bayesian analysis contends with the inherent degeneracy between
spin and charge, making the results more realistic. Simulations with higher SNRs
did not significantly alter these charge constraints, suggesting that including
additional QNMs in the ringdown model is crucial to obtaining more robust
limits.

\section{Conclusions\label{sec:Conslusions}}

In this study, our primary focus has been on constraining the charge of remnant
BHs using merger-ringdown GW data from GWTC-3 and simulated data from the ET. We
computed the posterior distributions for both charge and spin of the remnant
BHs. While our constraints from GWTC-3 data echoed previous findings, the
simulated ET data offered enhanced constraints. However, the overarching
challenge remains the degeneracy between charge and spin.

From the GWTC-3 dataset, we zeroed in on five events that exhibited trustworthy
ringdown signals. In our QNM analysis, we took into account both the 220 and 221
modes, deriving distributions for charge and spin of the resultant BHs. Notably,
these distributions are heavily influenced by the charge-spin degeneracy. This
is particularly evident in the case of GW191109, which displayed a pronounced
correlation between remnant charge and spin, resulting in a distinctive
belt-like distribution in the charge-spin plane. Upon employing a Kerr-based
prior informed by preceding IMR analyses, constraints such as \(Q<0.37\) emerged
from single events like GW200129 and GW200224, aligning closely with findings
from GWTC-2. Our subsequent evaluations of the Kerr hypothesis suggest that most
events exhibit charges consistent with zero, though the case for GW191109
remains somewhat ambiguous. Given the prevailing charge-spin degeneracy and the
current limitations in the sensitivity of detectors, all computed Bayes factors
between the KN BH and Kerr BH models are negative. {Although the Kerr BH model performs better in describing the five selected GW events, it is still insufficient to exclude the KN model.}

In our ET simulation, we generated synthetic ET data for BHs with varying
remnant charges. Given the elevated SNRs achievable with ET, we were able to
include 220, 221, and 330 modes in our analysis, thereby ameliorating the
charge-spin degeneracy issue to some extent. Consequently, the posterior
distributions for charge and spin improved substantially, manifesting
significantly reduced errors. However, complete elimination of charge-spin
degeneracy remains elusive, as evidenced by the persistence of belt-like
distributions in our results. Our simulation validated the efficacy of Bayes
factors in scrutinizing the Kerr hypothesis in the context of ET data.
Furthermore, the upper limit on charge obtained from the ``null test" improved
to \( Q<0.2 \) for \( a \approx 0.67 \) and \( Q<0.15 \) for \( a \approx 0.9 \)
with ET.

The crux of our investigation has been the persistent appearance of the
charge-spin degeneracy. This phenomenon stems from the similar dependencies of
currently utilized QNMs on both remnant charge and spin. The inclusion of
additional QNMs in future analyses could potentially alleviate this degeneracy,
thereby sharpening constraints on BH charge and facilitating more accurate
identification of charged remnant BHs. Given that ET provides the capability to
probe higher QNMs, future work will require expanded calculations on higher KN
QNMs. Finally, if there are some mechanism resulting in similar charge-to-mass
ratios for remnant BHs, one may also consider to perform event stacking in
order to obtain tighter constraints on the BH charge. Nevertheless, such
stacking relies on extra assumptions and is less general than the constraints
from single events as obtained in this work.

\begin{acknowledgments}

{We thank Gregorio Carullo, Harrison Siegel, and the anonymous referee for
helpful comments.} This work was supported by the National Natural Science
Foundation of China (11975027, 12247152, 11991053, 11721303), the China
Postdoctoral Science Foundation (2022TQ0011), the National SKA Program of China
(2020SKA0120300), the Max Planck Partner Group Program funded by the Max Planck
Society, and the High-performance Computing Platform of Peking University.  HTW
is supported by the Opening Foundation of TianQin Research Center. 

LIGO Laboratory and Advanced LIGO are funded by the United States National
Science Foundation (NSF) as well as the Science and Technology Facilities
Council (STFC) of the United Kingdom, the Max-Planck-Society (MPS), and the
State of Niedersachsen/Germany for support of the construction of Advanced LIGO
and construction and operation of the GEO600 detector.  Additional support for
Advanced LIGO was provided by the Australian Research Council.  Virgo is funded,
through the European Gravitational Observatory (EGO), by the French Centre
National de Recherche Scientifique (CNRS), the Italian Istituto Nazionale di
Fisica Nucleare (INFN) and the Dutch Nikhef, with contributions by institutions
from Belgium, Germany, Greece, Hungary, Ireland, Japan, Monaco, Poland,
Portugal, Spain.  KAGRA is supported by Ministry of Education, Culture, Sports,
Science and Technology (MEXT), Japan Society for the Promotion of Science (JSPS)
in Japan; National Research Foundation (NRF) and Ministry of Science and ICT
(MSIT) in Korea; Academia Sinica (AS) and National Science and Technology
Council (NSTC) in Taiwan of China.  

\end{acknowledgments}

\bibliography{refs}

\end{document}